\documentclass{article}
\usepackage{graphicx}
\begin{document}
\title{Synchronization and spindle oscillation in noisy integrate-and-fire-or-burst neurons with inhibitory coupling}
\author{Hidetsugu Sakaguchi and Shoko Tobiishi,\\
Department of Applied Science for Electronics and Materials,\\
 Interdisciplinary of Engineering Sciences,\\
 Kyushu University, Kasuga Fukuoka 816-8580, Japan}
\maketitle

We propose another integrate-and-fire model as a single neuron model.
We study a globally coupled noisy integrate-and-fire model with inhibitory interaction using the Fokker-Planck equation and the Langevin equation, and find a reentrant transition of oscillatory states. Intermittent time evolutions of neuron firing are found in strongly inhibited systems.  We propose another integrate-and-fire-or-burst model including the dynamics of the low-threshold Ca$^{2+}$ current based on the new integrate-and-fire model.  We study a globally coupled noisy integrate-and-fire-or-burst model with inhibitory interaction using the Fokker-Planck equation, and find bistability of the tonic mode and burst mode. Doubly periodic oscillation appears in a coupled system of two neuron assemblies, which is similar to the spindle oscillation in thalamic cells. 

\section{Introduction}
The limit cycle oscillation and the synchronization are found in various natural phenomema such as the firing of fireflies and the stick-slip motion of a fault in the earthquake. Coherent oscillation appears as the global synchronization of the coupled oscillators. The synchronization of oscillators is also considered to play important roles in neural information processing \cite{rf:1}.    
 Winfree and Kuramoto studied the global synchronization in general coupled oscillators including the phase oscillators \cite{rf:2,rf:3}.  The leaky-integrate-and-fire model (IF model) is one of the simplest models for a single neuron and often used to study dynamical behaviors of neural networks.  Each neuron receives an input via synaptic connections from other neurons. The neuron fires, when the membrane potential goes over a threshold, and sends out an impulse as an output to other neurons.   
Mirollo and Strogatz studied a globally coupled system of integrate-and-fire neurons, and showed that perfect synchronization occurs in a finite time \cite{rf:4}. Mutual synchronization is often observed among a large number of neurons with excitatory interaction. Inhibitory neurons can exhibit synchronization under some appropriate conditions \cite{rf:5,rf:6,rf:7}. However, the synchronization among inhibitory neurons is not so robust against some noises or heterogeneities, as was studied in several models in Refs.~[8] and [9]. On the other hand, the synchronization among thalamic neurons is known to be rather robust experimentally, although the synaptic coupling is inhibitory. The synchronous firing among the thalamic neurons generates characteristic brain waves called spindle waves in an early stage of sleep \cite{rf:10}. The reason of the robust synchronization  among inhibitory neurons is not well understood. In this paper, we will study the synchronization among noisy inhibitory neurons, using a new type of IF model.  

A single neuron may be affected by environmental noises, then, the model equation is generalized to a Langevin type equation such as a noisy integrate-and-fire model. 
In the noisy system, the coherent oscillation appears as an analogue of the phase transition in the statistical mechanics. 
Nykamp-Tranchina and Brunel-Hakim  showed that a population density approach using the Fokker-Planck equation is a useful method to study a large neural network of noisy integrate-and-fire neurons \cite{rf:11,rf:12,rf:13}. 
We studied a nonlocally coupled noisy integrate-and-fire model and found a traveling pulse state using the Fokker-Planck equation corresponding to the Langevin equation \cite{rf:14}.  The phase transition in a large population of noisy elements can be  studied with the Fokker-Planck equation in a clear manner, since the Fokker-Planck equation is a deterministic equation. In this paper, we apply the method to a large population of noisy integrate-and-fire neurons to clarify several phase transitions in some models of neural networks, although related subjects have been intensively studied with different methods and models \cite{rf:15,rf:16,rf:17}.   

The sleep spindle oscillations are a kind of brain waves, which appear in an early stage of sleep. The oscillations have a characteristic waveform of waxing-and-waning.  It is known that these oscillations are generated by the interaction between thalamic reticular (RE) and thalamocortical (TC) cells \cite{rf:10,rf:18,rf:19}. The interactions from RE cells to TC cells are inhibitory and the interactions from TC cells to RE cells are excitatory, the mutual interaction among RE cells is inhibitory, and there is no mutual interaction among TC cells. The 
low-threshold Ca$^{2+}$ current plays an important role for the excitation of the thalamic cells. The integrate-and-fire-or-burst model (IFB model) is a generalized model of the IF model, which takes the dynamics of the low-threshold Ca$^{2+}$ current into consideration~\cite{rf:18}. The dynamics of the IFB model under external stimuli was studied in Ref.~[19].
We will propose a new type of IFB model including the dynamics for the low-threshold Ca$^{2+}$ current based on the new IF model. We will study a globally coupled noisy IFB model with inhibitory interaction as a neural network model of the RE cells, and finally study a coupled systems of two types of IFB neuron assemblies as a coupled model of the RE cells and TC cells.   

We will investigate the coupled noisy integrate-and-fire models stepwise. 
In \S 2, we propose a new type of integrate-and-fire (IF) model with inhibitory coupling and show the synchronization between two neurons. In \S 3, we investigate a globally coupled noisy integrate-and-fire model using the Fokker-Planck equation. In \S 4, we propose a new type of integrate-and-fire-or-burst (IFB) model by combining the new type of integrate-and-fire model with the dynamics of the low-threshold Ca$^{2+}$ current, and investigate the global oscillation in the globally coupled noisy IFB model. In \S 5, we investigate a coupled system of two neuron assemblies and show doubly periodic oscillation. 

\section{Synchronization of two integrate-and-fire neurons with inhibitory coupling}
In the usual integrate-and-fire model, the membrane potential $v$ is reset to a certain value $V_R\sim -50 $ mV, just after the membrane potential reaches a threshold value, which is assumed to be about $-35\sim -45$ mV. In more realistic models, the neuron is excited after the membrane potential goes over the threshold, the membrane potential takes positive values for a while, and then the membrane potential returns to a negative value around the resting potential.         
Taking the dynamics in the excited state into consideration, we propose another type of simple integrate-and-fire model. Our new IF model keeps the most favorable point of the IF model that the firing of a neuron can be described using only one variable, in contrast to the Hodgkin-Huxley model including four variables~\cite{rf:20} or the FitzHugh-Nagumo model including two variables~\cite{rf:21}. 
 
We study the synchronization of two neurons with inhibitory coupling, using the new IF model. The model equation is written as 
\begin{eqnarray}
C\frac{dv_i}{dt}&=&-\alpha(v_i-V_0)+I_0+I_s, \;\;{\rm for}\;\;v_i<V_T,\nonumber\\
C\frac{dv_i}{dt}&=&-\alpha(v_i-V_0^{\prime})+I_s, \;\;{\rm for}\;\;v_i>V_T,
\end{eqnarray}
where $v_i$ denotes the membrane potential for the $i$th cell, $V_0=-65$ denotes the resting potential, $V_0^{\prime}$ is another parameter which controls the dynamics in the excited state $v_i>V_T$ and is fixed to be $V_0^{\prime}=35$ in this paper, $I_0$ is a constant external stimulus current, $I_s$ is a synaptic current, and $C=2$ and $\alpha=0.035$ are constants. The units of $v$ and $t$ are mV and ms. When the membrane potential exceeds the threshold $V_T=-35$, the cell is excited. In the usual integrate-and-fire model, $v_i$ is reset to a potential $V_R=-50$ just after the firing.  However, in our generalized model, the membrane potential jumps to a potential $V_1>0$ ($V_1=60$ is assumed in this paper.)  and then it tends to decay toward $V_0^{\prime}$ if $I_s=0$, obeying the second equation of Eq.~(1). When the membrane potential reaches a second threshold $V_2$, $v_i$ is reset to $V_R$.  The two kinds of time evolutions correspond to the slow dynamics on the two branches of nullclines in the excitable equations such as FitzHugh-Nagumo model~\cite{rf:21} and the McKean model~\cite{rf:22}. 

\begin{figure}[htb]
\begin{center}
\includegraphics[width=11.cm]{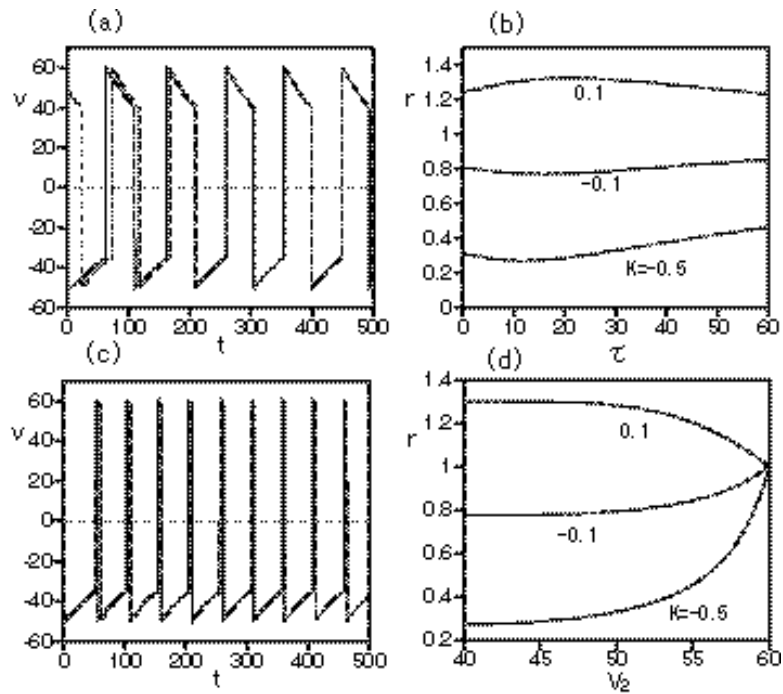}
\end{center}
\caption{(a) Two time evolutions by Eq.~(1) starting from $v_1=-50$ (solid line), $v_2=50$ (dashed line) and $I_s=-0.9973$ for $I_0=1.5,\,K=-0.5,\,V_2=40$ and $\tau=10$. (b) Stability exponent $r$ as a function of $\tau$ for  $K=0.1,-0.1$ and $-0.5$ at $I_0=1.5$ and $\tau=10$. (c) Two time evolutions by Eq.~(1) starting from $v_1=-48$ (solid line) and $v_2=59.9$ (dashed line) and $I_s=-0.9973$ for $I_0=1.5,\,K=-0.5,\,V_2=58$ and $\tau=10$. (d) Stability exponent $r$ as a function of $V_2$ for $K=0.1,-0.1$ and $-0.5$ at $I_0=1.5$ and $\tau=10$.}
\label{fig:1}
\end{figure}
There are several model equations which describe the dynamics of the synaptic current $I_s(t)$~\cite{rf:23}.  In this paper, we use a simple model including a response time. The synaptic current for a coupled system of two neurons is assumed to obey  
\begin{equation}
\tau\frac{dI_s}{dt}=-(I_s-K/2\sum_{j=1}^2\theta(v_j)),
\end{equation}
where $\tau$ is a time constant, $K<0$ is an inhibitory coupling constant, and $\theta(x)$ is the Heaviside step function. If the cell is excited ($v_j>0$), the inhibitory synaptic current $I_s<0$ is induced. This type of simple synaptic coupling is often used in some theoretical models \cite{rf:24}.
  
Figure 1(a) displays the time evolution of $v_1$ and $v_2$ starting from $v_1(0)=-50,\,v_2(0)=50$ and $I_s(0)=-0.9973$ at $I_0=1.5,K=-0.5, V_2=40$ and $\tau=10$. The two oscillatory time evolutions approach each other, which means that the mutual synchronization is attained. 
The solution of the synchronized oscillation can be explicitly solved, because the model equation is a piecewise linear equation. The stability of the synchronized motion can be studied by the linearized equation of Eqs.~(1) and (2). The deviation $\delta v=v_1-v_2$ obeys a simple equation: $Cd\delta v/dt=-\alpha\delta v$. The deviation $\delta v$ changes discontinuously at the moment $t_1$ of the first threshold satisfying $v_1(t_1)=V_T$, as $\delta v(t_{1+})=(dv/dt)_{v_1=V_1}/(dv/dt)_{v_1=V_T}\delta v(t_{1-})$, and at the moment $t_2$ of the second threshold satisfying $v_1(t_2)=V_2$, as $\delta v(t_{2+})=(dv/dt)_{v_1=V_R}/(dv/dt)_{v_1=V_2}\delta v(t_{2-})$. Therefore, the deviation $\delta v(t)$ grows $\delta v(t+T)=r \delta v(t)$ after a period $T$ of the synchronized motion, where the stability exponent $r$ is expressed as 
\begin{equation}
r=\frac{(dv/dt)_{v_1=V_1}}{(dv/dt)_{v_1=V_T}}\frac{(dv/dt)_{v_1=V_R}}{(dv/dt)_{v_1=V_2}}e^{-\alpha/C\cdot T}.
\end{equation}
Figure 1(b) displays the stability exponent $r$ as a function of $\tau$ for $I_0=1.5,K=-0.5$ and $\tau=10$.  If the stability exponent $r$ is smaller than 1, the synchronized motion is stable.  As $K<0$ and $|K|$ is larger, the synchronized motion becomes more stable. There is an optimum response time $\tau$ where $r$ takes a minimum value, although the synchronized motion is stable even for very small $\tau$ for $K<0$. In the case of excitatory coupling $K>0$, $r$ is larger than 1 and the synchronized motion is unstable. That is, the mutual synchronization is attained in the case of inhibitory coupling and the synchronized motion becomes more stable, as the inhibitory coupling is stronger.  
 
If the second threshold $V_2$ approaches $V_1$, the duration time of the excitation becomes smaller, and in the limit of $V_2=V_1$, the duration time of the excitation becomes 0 and our model approaches the original integrate-and-fire model.  Figure 1(c) displays the mutual synchronization of two neurons starting from $v_1=-48$ (solid line) and $v_2=59.9$ (dashed line) and $I_s=-0.9973$ at $I_0=1.5,\,K=-0.5,\,V_2=58$ and $\tau=10$. The duration time of the excited state is much shorter than the case of Fig.~1(a) owing to the smallness of $|V_1-V_2|$, however, the mutual synchronization is still attained.  
Figure 1(d) displays the stability exponent $r$ of the synchronized state as a function of $V_2$ for $K=-0.5,\,-0.1$ and 0.1 at $I_0=1.5$ and $\tau=10$. 
As $V_2$ approaches $V_1=60$, the stability exponent approaches 1, because the duration time of the excited state becomes shorter, and then the synaptic interaction becomes weaker effectively. In the following numerical simulations, we use the parameter value of $V_2=40$ (except in Fig.~4(b)), because mutual synchronization occurs fairly strongly, although the duration time of the excitation might be too long as a typical neuron.  

\section{Global synchronization in a large population of noisy integrate-and-fire neurons}
Next, we consider a large population of integrate-and-fire neurons with inhibitory coupling. The total number of the neurons is assumed to be $N$. The elemental dynamics for each neuron obeys Eq.~(1).  The synaptic current is expressed  as 
\[
\tau\frac{dI_s}{dt}=-(I_s-K/N\sum_{j=1}^N\theta(v_j)).
\]
A completely synchronized state where $v_i(t)=v(t)$ for any $i$ is a special solution to the globally coupled system, and the stability exponent $r$ of the completely synchronized state is calculated from the linear growth rate of $\delta v_i(t)=v_i(t)-v(t)$. The stability exponent $r$ takes the same value as that in a coupled system of two neurons, because $K/N\sum_{j=1}^N\theta(v_j)$ and $I_s(t)$ take the same value as the case of two neurons.  That is, the completely synchronized state is stable for the globally coupled system with inhibitory interaction without noise terms.  

If a noise term is added to Eq.~(1), the model equation is expressed as a Langevin equation:  
\begin{eqnarray}
C\frac{dv_i}{dt}&=&-\alpha(v_i-V_0)+I_0+I_s+C\xi_i(t), \;\;{\rm for}\;\;v_i<V_T,\nonumber\\
C\frac{dv_i}{dt}&=&-\alpha(v_i-V_0^{\prime})+I_s+C\xi_i(t), \;\;{\rm for}\;\;v_i>V_T,\nonumber\\
\tau\frac{dI_s}{dt}&=&-\{I_s-\frac{K}{N}\sum_{j=1}^N\theta(v_j)\},
\end{eqnarray}
where $\xi_i(t)$ denotes Gaussian white noise which satisfies $\langle \xi_i(t)\xi_j(t^{\prime})\rangle=2D\delta_{i,j}\delta(t-t^{\prime})$. 
The corresponding Fokker-Planck equation for $N\rightarrow \infty$ is written as \begin{eqnarray}
\frac{\partial P}{\partial t}&=&-\frac{\partial}{\partial x}\{f(x,I_0,I_s)P(x)\}+D\frac{\partial^2P}{\partial x^2}+\delta(x-V_1)J_1(t)+\delta(x-V_R)J_2(t),
\nonumber\\
\tau\frac{dI_s}{dt}&=&-(I_s-K\int_{0}^{\infty}P(x)dx),
\end{eqnarray}
where $P(x,t)$ denotes the probability density that the membrane potential $v$  takes a value $x$, $J_1(t)=-D(\partial P/\partial x)_{x=V_T}$ and $J_2(t)=-D(\partial P/\partial x)_{x=V_2}$. The function $f(x,I_0,I_s)$ denotes a simplified form of the deterministic part of the elemental dynamics  for each neuron, that is,  the first two equations in Eq.~(4) are rewritten as $dv_i/dt=f(v_i,I_0,I_s)+\xi_i(t)$ for the sake of simplicity. 
The terms including the $\delta$-function represent the two jump processes from $v=V_T$ to $V_1$ and from $V_2$ to $V_R$. 
The probability density $P(x)$ is 0 at $x=V_T$ and $V_2$.  The integral $\int_{0}^{\infty}P(x)dx$ denotes the fraction of the excited neurons, which is equal to $\sum_{j=1}^N\theta(v_j)/N$. We have performed direct numerical simulation of Eq.~(5) with the Euler method with $\Delta x=0.1$ and $\Delta t=0.005$. 
Similar direct numerical simulation was performed for the usual integrate-and-fire neurons with only one threshold in Ref.~[14].

\begin{figure}[htb]
\begin{center}
\includegraphics[width=13cm]{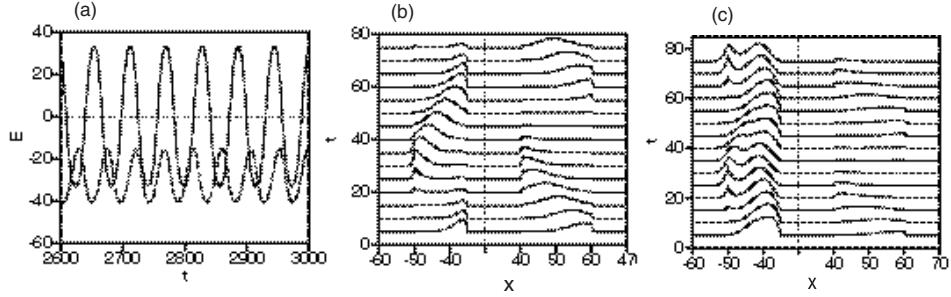}
\end{center}
\caption{(a) Time evolution of the average membrane potential $E(t)$ by Eq.~(5) at $K=-2$ (solid line) and $K=-10$ (dashed line) for $I_0=2.5,D=0.2,V_2=40$ and $\tau=20$. 
(b) Time evolution of the probability density $P(x,t)$ at $K=-2$. Since $P(x,t)$ is always 0 in $-30<x<30$, the region is not plotted.  (c) Time evolution of the probability density $P(x,t)$ at $K=-10$.}
\label{fig:2}
\end{figure}

\begin{figure}[htb]
\begin{center}
\includegraphics[width=10cm]{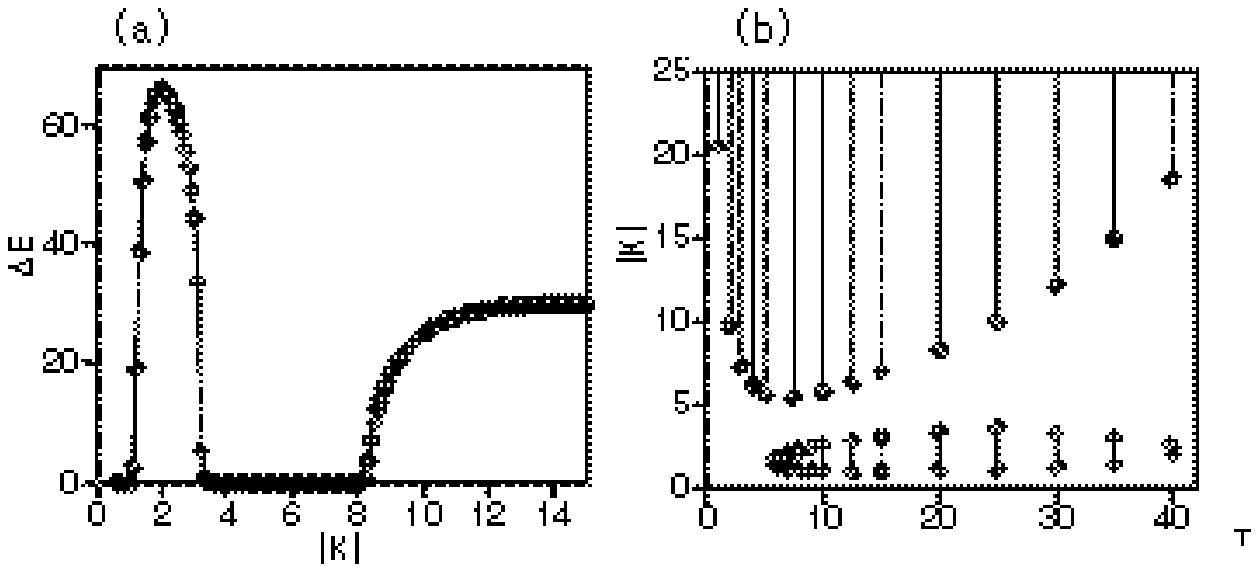}
\end{center}
\caption{(a) Oscillation amplitude as a function of $|K|$ for $I_0=2.5,D=0.2$ and $\tau=20$. (b) Parameter region where the global synchronization appears in the parameter space of $(\tau,|K|)$ for $I_0=2.5$ and $D=0.2$. The coupling constant $K$ is negative, and the same parameter values $I_0$ and $D$ are used for (a) and (b)}
\label{fig:3}
\end{figure}
We have calculated the average value of the membrane potential $E=\int_{-\infty}^{\infty} P(x,t)xdx$. Figure 2(a) displays time evolutions of $E(t)$ at $K=-2$ (solid line) and $-10$ (dashed line) for $I_0=2.5,\,D=0.2$ and $\tau=20$. Clear limit cycle oscillations of the average membrane potential are observed, which implies the synchronization of firings in noisy environment. At $K=-2$, the average membrane potential goes up to $E\sim 35$, but at $K=-10$, the average membrane potential is always below 0. Figures 2(b) and (c) display time evolutions of the probability density $P(x,t)$ at (b) $K=-2$ and (c) $K=-10$ for $I_0=2.5,\,D=0.2$ and $\tau=20$. The probability density is oscillating across the threshold $x=V_T$ as a whole at $K=-2$, but only a fraction of probability density goes up to the positive region $x>0$ at $K=-10$, remaining most part of probability density below the threshold.    
That is, almost all neurons fire synchronously at $K=-2$, but only a fraction of neurons fire synchronously at $K=-10$, which will be clearly shown by the corresponding Langevin simulation in Fig.~4. 

Figure 3(a) displays the peak-peak amplitude $\Delta E=E_{max}-E_{min}$ of the oscillation of the average membrane potential $E(t)$ for $I_0=2.5,\,\tau=20$ and $D=0.2$ as a function of the coupling constant $|K|=-K$. As is seen in Fig.~3(a), the global synchronization occurs for  $1.1\le |K|\le 3.4$ and $|K|\ge 8.3$. The transitions are supercritical Hopf bifurcations. The global synchronization reappears in a parameter region of smaller $|K|$. That is, a reentrant transition of the global synchronization occurs in this noisy system. Figure 3(b) shows a parameter region where the global synchronization is observed in a parameter space of $\tau$ and $|K|$ for $I_0=2.5$ and $D=0.2$.  There are two parameter regions where the global oscillation occurs. In the parameter region of smaller $|K|$, almost all neurons fire synchronously. The parameter region of the smaller $|K|$ is bounded between $6<\tau<40$. This might be related to a fact that the stability exponent $r$ is the smallest near $\tau=15$ for $I_0=1.5$ as shown in Fig.~1(b), that is, the finite response time facilitates the global synchronization.   
In the parameter region of larger $|K|$, a fraction of neurons fire synchronously, but the average membrane potential is lower than 0. The strong inhibition with large $K$ inhibits the firing of all neurons.  Neurons which fire earlier, inhibit the firing of the other neurons. Which neuron fires is not determined but change randomly in our noisy system, in contrast to a clustered state found in some deterministic systems \cite{rf:25}.  
That is, the global oscillation appears, but the globally synchronous firing does not occur in the region of large $K$.  

We have shown numerical results for the Fokker-Planck equation. 
The Fokker-Planck equation is favorable to investigate the global oscillation clearly, since the equation is deterministic.  However, the time evolution of individual neuron cannot be seen. The direct Langevin simulation can show the time evolutions of each neuron.
\begin{figure}[htb]
\begin{center}
\includegraphics[width=8cm]{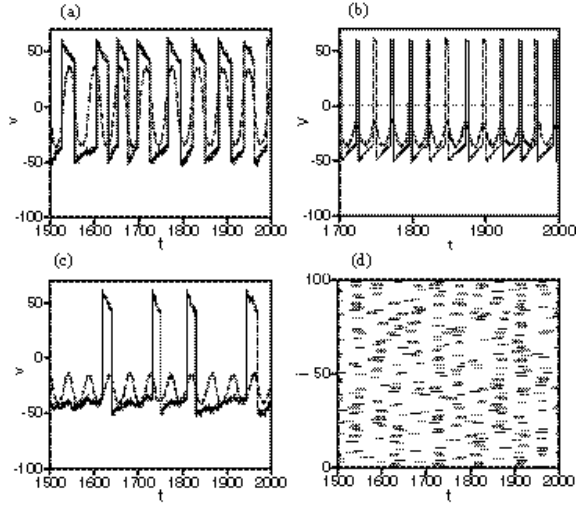}
\end{center}
\caption{Time evolutions of $v_i$ (solid line) and the average $\langle v\rangle$ (dashed line) in the Langevin simulation of Eq.~(4) for a globally coupled noisy IF model  at  (a) $K=-2,D=0.2,V_2=40$, (b) $K=-2,D=0.02,V_2=58$ and (c) $K=-10,D=0.2,V_2=40$ for $I_0=2.5,\tau=20$ and $N=2000$. (d) Raster plot at $I_0=2.5,K=-10,D=0.2,V_2=40,\tau=20$ and $N=2000$ which corresponds to (c), in which a dot is plotted at time $t=n$ ($n$ is an integer) and the position $i$ satisfying $v_i(t=n)>0$ for $1\le i \le 100$.}
\label{fig:4}
\end{figure}
The difference of the two types of global oscillation can be clearly seen in the Langevin simulation of Eq.~(4). We have performed a direct Langevin simulation for $N=2000$ neurons, and got time evolutions of membrane potentials $v_i$'s for all neurons and the average value of the membrane potentials.
We have performed some numerical simulations by changing parameter values of $D$, $V_2$ and $K$.     
Figures 4(a), (b) and (c) display the time evolutions of $v_{i=1}(t)$ of the first neuron (solid line) and the average $\langle v\rangle=\sum_{j=1}^Nv_j/N$ (dashed line) for $I_0=2.5,\tau=20$ and $N=2000$. Figure 4(a) displays the time evolutions  at $K=-2,D=0.2$ and $V_2=40$, which corresponds to the solid line in Fig.~2(a). The firing of each neuron is almost synchronized to the average activity $\langle v\rangle$. The timing of the firing is slightly different for each neuron owing to the external noise, that is, the phase of the oscillator is randomized by the external noise. 
The average value $\langle v\rangle$ exhibits limit cycle oscillation, which is almost equal to $E(t)$ in Fig.~2(a). The global synchronization may be interpreted by the concept of the phase synchronization as studied in Ref.~[3].
Figure 4(b) displays the time evolution at $K=-2,D=0.02$ and $V_2=58$. At $V_2=58$, the duration time of the excited state is much shorter, and the effective interaction becomes weak as shown in Fig.~1(c) and (d), but almost synchronized firing is  seen for the smaller noise strength $D=0.02$. For the same parameter values $K=-2$ and $D=0.2$ as Fig.~4(a), the global synchronization was not observed at $V_2=58$, because the timing of the firing is randomly distributed. Inhibitory neurons can exhibit the global synchronization even if the duration time of the excited state is short, but the stability becomes weaker.    
 Figure 4(c) displays the time evolutions at $K=-10,D=0.2$ and $V_2=40$, which corresponds to the dashed line in Fig.~2(a). The average value $\langle v\rangle$ exhibits limit cycle oscillation, which is close to $E(t)$ in Fig.~2(a).  The activity $v_i(t)$ of each neuron shows that the firing occurs intermittently at some peaks of $\langle v\rangle$. A fraction of neurons fire near the peak of $\langle v\rangle$, and the other neurons are inhibited to fire by the strong global inhibition by the fired neurons.  Which neurons are selected for firing is not determined but changes randomly. As a result, the intermittent time evolution appears as shown in Fig.~4(c).  The random but somewhat synchronized  firing is seen in the raster plot of Fig.~4(d) at the same parameter values as in Fig.~4(c), where firing sequences of neurons for $1\le i\le 100$ are denoted by dot patterns.
Intermittent firing was found also in a numerical simulation by Wang et al.\cite{rf:16}, but their model is deterministic and the coupling strengths among neurons are randomly distributed. Similar intermittent behaviors are observed also experimentally~\cite{rf:26}.

\section{Bistability in a large population of the integrate-and-fire-or-burst neurons}
The integrate-and-fire model is a simpler model than  realistic model equations such as the Hodgkin-Huxley equation~\cite{rf:20}, in that the detailed dynamics of ion channels are neglected. Specific ion channels play an important role in the firing process of specific neurons.  The low-threshold Ca$^{2+}$ current is important for thalamic neurons. 
The integrate-and-fire-or-burst model was proposed as a simple model to describe the dynamics of the membrane potential of thalamic cells \cite{rf:18}. It is a coupled equation of the integrate-and-fire model and the dynamics of the low-threshold Ca$^{2+}$ current $I_T(t)$. The current $I_T(t)$ is given by $I_T(t)=g_Th(t)(V_H-v(t))\theta(v(t)-V_h)$, where the parameters are given as $g_T=0.07, V_H=120$ and $V_h=-60$, and $h(t)$ is a slow variable determined by the membrane potential $v$.  In the integrate-and-fire-or-burst model, the slow variable $h(t)$ ($0\le h\le 1$) is assumed to obey 
\[
\tau_h(v)\frac{dh}{dt}=-h+\theta(V_h-v(t)),
\]
where $\tau_h=20$ for $v>V_h$ and $\tau_h=100$ for $v<V_h$ \cite{rf:18,rf:19}. 
The low-threshold current $I_T$ is zero, if the membrane potential is constant.
The current $I_T$ flows in the neuron for a short time $\sim\tau_h$, after the membrane potential $v$ goes over the threshold $V_h$ from below. 
The integrate-and-fire-or-burst model is a coupled equation of the standard integrate-and-fire model and the equation of $h(t)$. We study a new integrate-and-fire-or-burst (IFB) model, which is a coupled equation of the new IF model Eq.~(1) and the equation of $h(t)$. We investigate a globally coupled noisy IFB model, which is expressed as 
\begin{eqnarray}
C\frac{dv_i}{dt}&=&-\alpha(v_i-V_0)+I_0+I_s+I_{Ti}+C\xi_i(t), \;\;{\rm for}\;\;v_i<V_T,\nonumber\\
C\frac{dv_i}{dt}&=&-\alpha(v_i-V_0^{\prime})+I_s+I_{Ti}+C\xi_i(t), \;\;{\rm for}\;\;v_i>V_T,\nonumber\\
\tau_h\frac{dh_i}{dt}&=&-h_i+\theta(V_h-v_i(t)),\;\;I_{Ti}(t)=g_Th_i(t)(V_H-v_i(t))\theta(v_i(t)-V_h),\nonumber\\
\tau\frac{dI_s}{dt}&=&-\{I_s-\frac{K}{N}\sum_{j=1}^N\theta(v_j)\}.
\end{eqnarray}

The slow variable $h_i(t)$ is another stochastic variable, because $v_i(t)$ is a stochastic variable. Therefore, the corresponding Fokker-Planck equation takes a form of a two-dimensional partial differential equation:   
\begin{eqnarray}
\frac{\partial P}{\partial t}&=&-\frac{\partial}{\partial x}\{f(x,I_0,I_s+I_T)\}P(x)-\frac{\partial}{\partial y}[\{(-y+\theta(V_h-x)/\tau_h(x)\}P]\nonumber\\
&+&D\frac{\partial^2P}{\partial x^2}+\delta(x-V_1)J_1(t,y)+\delta(x-V_R)J_2(t,y),
 \nonumber\\
\tau\frac{dI_s}{dt}&=&-(I_s-K\int_{0}^{\infty}\int_0^1P(x)dxdy),
\end{eqnarray}
where $P(x,y,t)$ is the probability density that the membrane potential $v$ takes a value $x$ and the slow variable $h$ takes a value $y$, $f(x,I_0,I_s+I_T)$ denotes the deterministic part of the elemental dynamics of the membrane potential, $I_T(x,y)=g_Ty(V_H-x)\theta(x-V_h)$, and $J_1(t,y)=-D(\partial P/\partial x)_{x=V_T}$ and $J_2(t,y)=-D(\partial P/\partial x)_{x=V_2}$. Since $h(t)$ does not receive a random force, there is no diffusion term with respect to $y$ in Eq.~(7). The deterministic forces to the membrane potential $v(t)$ and the variable $h(t)$ appear in the first two drift terms on the right-hand side of Eq.~(7).  
We have performed numerical simulations of the two-dimensional partial differential equation (7) using the Euler method with $\Delta x=0.1,\,\Delta t=0.005$ and $\Delta y=0.05$.  

\begin{figure}[htb]
\begin{center}
\includegraphics[width=9cm]{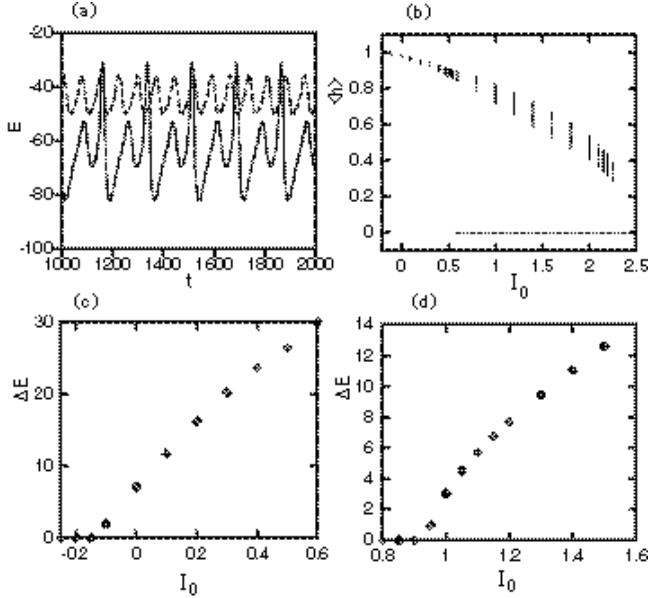}
\end{center}
\caption{(a) Two time evolutions of the average membrane potential $E(t)$ by Eq.~(7) for $I_0=1.6,K=-40,D=0.2$ and $\tau=20$. The initial conditions for the two time evolutions are different. (b) The range of the temporal variation of the average value:$\langle h\rangle$ as a function of $I_0$ for $K=-40,D=0.2$ and $\tau=20$. The solid line denotes the burst mode and the dashed line denotes the tonic mode. (c) Oscillation amplitude of $E(t)$ as a function of $I_0$ for the burst mode at $K=-40$ and $D=0.2$. (d) Oscillation amplitude of $E(t)$ as a function of $I_0$ for the tonic mode at $K=-40$ and $D=0.2$.}
\label{fig:5}
\end{figure}
Figure 5(a) displays two time evolutions of $E(t)=\int_{-\infty}^{\infty}\int_0^1P(x,y)xdxdy$ for $I_0=1.6,K=-40,D=0.2$ and $\tau=20$. The initial conditions are different for the two time evolutions.  That is, the two kinds of stable oscillatory states have appeared for the same parameters. It implies the bistability.  The average membrane potential is always larger than $V_h$ in the dashed line, so $h(t)$ is almost 0. The low-threshold Ca$^{2+}$ current does not play a role, and the frequency of the oscillation is relatively high. This is called the tonic mode. On the other hand, the average membrane potential $E(t)$  oscillates around $V_h=-60$ in the solid line. The low-threshold Ca$^{2+}$ current plays an important role in this mode. The low-threshold current $I_T$ depolarizes the IFB neuron and the firing is facilitated. If many neurons are fired, the strong inhibitory coupling hyperpolarizes each neuron and the membrane potential goes down below $V_h$, and then is rebound to go over the threshold $V_h$ again.  The low-threshold Ca$^{2+}$ current facilitates the oscillation of the firing, and therefore the amplitude of oscillation is relatively large and the frequency of oscillation is relatively low in this mode. This mode is called the burst mode or the rebound mode.  
The tonic mode and the burst mode are bistable at the parameters in Fig.~5(a) in our model. Figure 5(b) displays the range of the temporal oscillation of the average $\langle h(t)\rangle=\int_{-\infty}^{\infty}\int_0^1 P(x,y,t)ydxdy$ as a function of $I_0$ for $K=-40,D=0.2$ and $\tau=20$. The finite width of the temporal variation of  $\langle h(t)\rangle$ at a certain $I_0$ implies the limit cycle oscillation of $\langle h(t)\rangle$. The upper branch represents the burst mode and the lower branch ($\langle h\rangle=0$) represents the tonic mode. (We have distinguished the two modes in this paper from a viewpoint that the low-threshold Ca$^{2+}$ current plays an important role or not, that is,  the average value of $h(t)$ is finite or almost zero.)
The bistability occurs for $0.55\le I_0\le 2.3$. 
Figure 5(c) displays the peak-peak amplitude $\Delta E$ of the burst oscillation as a function of $I_0$. The supercritical Hopf bifurcation is seen at $I_0\sim -0.15$. Near the threshold, the oscillation is sinusoidal, but a period doubling bifurcation occurs for the burst mode for $I_0>0.7$. The burst-mode oscillation shown by the solid line in Fig.~5(a) displays the time evolution of $E(t)$ after the period doubling bifurcation. 
In the tonic mode, the Hopf bifurcation occurs near $I_0=0.95$, as shown in Fig.~5(d).  Figure 5(c) and (d) show that the oscillation amplitude in the burst mode is relatively large compared to the tonic mode.

\section{Spindle oscillation in a coupled system of two neuron assemblies}
We investigate a coupled system of two neuron assemblies in this section, as a model of the thalamic neural network. 
There are two kinds of neurons: thalamic reticular (RE) cells and thalamocortical (TC) cells in the thalamus.  The low threshold Ca$^{2+}$ current plays an important role in both types of thalamic cells. The interaction among RE cells is inhibitory. In the previous section, we have investigated the dynamical behaviors of a globally coupled system of one kind of inhibitory IFB neurons, which corresponds to an assembly of RE cells alone.  
It is known that the interaction between the RE cells and the TC cells is important for generation of spindle waves, which are characteristic brain waves in the second stage of sleep. We will study a coupled system of two neuron assemblies corresponding to RE cells and TC cells. It is known that the interaction from the RE cells to the TC cells is inhibitory and the interaction from the TC cells to the RE cells is excitatory, and there is no mutual interaction among the TC cells.  We investigate a Fokker-Planck equation for the coupled system of two neuron assemblies composed of noisy IFB neurons.
The Fokker-Planck equation for the coupled neuron assemblies is expressed as  
\begin{eqnarray}
\frac{\partial P_1}{\partial t}&=&-\frac{\partial}{\partial x}
\{f(x,I_{01},I_{s1}+I_{s2}+I_T)P_1\}-\frac{\partial}{\partial y}[\{(-y+\theta(V_{h1}-x))/\tau_h\}P_1]\nonumber\\&+&D\frac{\partial^2P_1}{\partial x^2}+\delta(x-V_1)J_{11}(t,y)+\delta(x-V_R)J_{12}(t,y),\nonumber\\
\frac{\partial P_2}{\partial t}&=&-\frac{\partial}{\partial x}\{f(x,I_{02},I_{s3}+I_T)P_2\}-\frac{\partial}{\partial y}[\{(-y+\theta(V_{h2}-x))/\tau_h\}P_2\}]\nonumber\\&+&D\frac{\partial^2P_2}{\partial x^2}+\delta(x-V_1)J_{21}(t,y)+\delta(x-V_R)J_{22}(t,y),\nonumber\\
\tau_1\frac{dI_{s1}}{dt}&=&-(I_{s1}-K_1\int_{0}^{\infty}\int_0^1P_1(x)dxdy),\nonumber\\
\tau_2\frac{dI_{s2}}{dt}&=&-(I_{s2}-K_2\int_{0}^{\infty}\int_0^1P_2(x)dxdy),\nonumber\\
\tau_3\frac{dI_{s3}}{dt}&=&-(I_{s3}-K_3\int_{0}^{\infty}\int_0^1P_1(x)dxdy),
\end{eqnarray}
\begin{figure}[htb]
\begin{center}
\includegraphics[width=10cm]{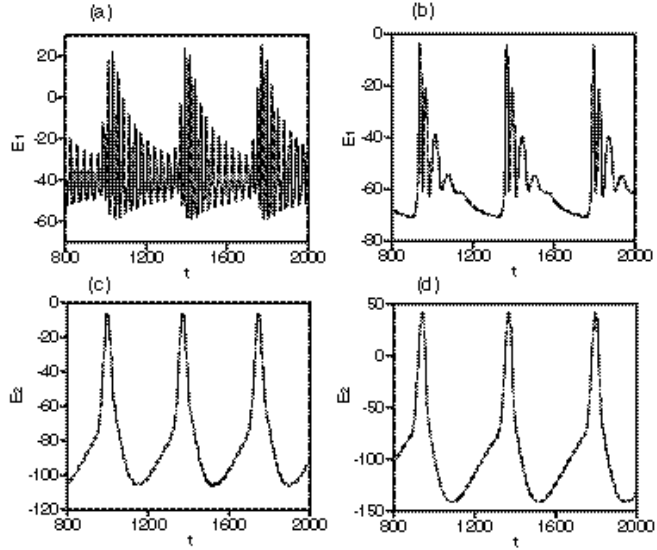}
\end{center}
\caption{Time evolutions of the average membrane potential $E_1(t)$ of the RE neurons in Eq.~(8) at (a) $I_{01}=I_{02}=2.2$ and (b)$I_{01}=-0.25,\,I_{02}=1$, and the corresponding average membrane potential $E_2(t)$ of the TC neurons at (c)$I_{01}=I_{02}=2.2$ and (d)$I_{01}=-0.25,\,I_{02}=1$.}
\label{fig:6}
\end{figure}
where $P_1(x,y,t)$ and $P_2(x,y,t)$ are the probability densities for the two neuron assemblies, $J_{11}(t,y)=-D(\partial P_1/\partial x)_{x=V_T}$,$J_{12}(t,y)=-D(\partial P_1/\partial x)_{x=V_2}$,$J_{21}(t,y)=-D(\partial P_2/\partial x)_{x=V_T}$ and $J_{22}(t,y)=-D(\partial P_2/\partial x)_{x=V_2}$. 
Our model is a mathematical model and may not describe the dynamics of realistic thalamic cells well, but we call the first neuron assembly the RE neurons and the second neuron assembly the TC neurons, for the sake of simplicity.  
We assume the sign of the three coupling constants as $K_1<0$, $K_2>0$ and $K_3<0$, taking the thalamic neural network into consideration. The synaptic current $I_{s1}<0$ represents the inhibitory coupling among the RE neurons, $I_{s2}>0$ represents the excitatory coupling from the TC neurons to the RE neurons, and $I_{s3}<0$ represents the inhibitory coupling from the RE neurons to the TC neurons. 
The external inputs to the RE and TC neurons are respectively denoted by $I_{01}$ and $I_{02}$. The response time $\tau_1$ for the RE neurons is assumed to be smaller than the response times $\tau_2$ and $\tau_3$ for the interaction between the RE and TC neurons, since the distance between the RE neurons and TC neurons is larger than the distance among the RE neurons. The threshold values of the low-threshold Ca$^{2+}$ current for the RE neuron and the TC neuron are respectively assumed to be $V_{h1}=-60$ and $V_{h2}=-70$ \cite{rf:19}.
This model equation is rather complicated and includes many parameters. We have performed numerical simulations of Eq.~(8) for $\tau_1=5,\,\tau_2=\tau_3=100,\,K_1=K_3=-30,\,K_2=30,\,D=0.2$ as an example. Figure 6 displays the time evolutions of the average membrane potential $E_1(t)=\int_{-\infty}^{\infty} \int_0^1P_1(x,y,t),xdxdy$ of the RE neurons for (a) $I_{01}=I_{02}=2.2$ and (b) $I_{01}=-0.25,\,I_{02}=1$, and the average membrane potential $E_2(t)$ of the TC neurons for for (c) $I_{01}=I_{02}=2.2$ and (d) $I_{01}=-0.25,\,I_{02}=1$. Doubly-periodic oscillations are seen in the time evolutions of $E_1(t)$.  The long-term oscillatory motion is caused by the mutual interaction between the RE and TC neurons. The fast oscillation appears owing to the synchronization among the RE neurons, which was shown in the previous section. The two kinds of oscillations are overlapped and the doubly-periodic motion appears in the time evolution of $E_1(t)$.  The mechanism of the doubly periodic motion is as follows. The fast oscillation of the RE neurons repeats for a while. The TC neurons are inhibited strongly below the $V_{h2}$ by the repetition of the fast oscillations of the RE neurons. 
The firing of the RE neurons becomes weak, as the input from the TC neurons becomes small. The TC  neurons are rebound after the strong inhibition, and the membrane potential of the TC neurons goes over $V_{h2}$.  The low threshold Ca$^{2+}$ current flows into the TC neurons and the TC neurons fire. Then, the RE neurons are excited by the firing of the TC neurons. The long period is determined by the slow rebound time.  The waxing-and-waning motion of the fast oscillation is seen clearly in Fig.~6(b). The fast oscillations repeat several times and then disappear. In the repetition of the fast oscillations, the period increases gradually, the amplitude of the fast oscillation is decreasing. This type of motion is characteristic of the spindle oscillation in the thalamic networks~\cite{rf:27}. 

In our model, the doubly periodic oscillation is caused by the interaction between the RE neurons and TC neurons, but there are other neuron models where the slow oscillation is caused by slow dynamics of some ion channels~\cite{rf:28,rf:29}. In their models, the doubly periodic behavior can be generated even in one neuron.  

\begin{figure}[htb]
\begin{center}
\includegraphics[width=6cm]{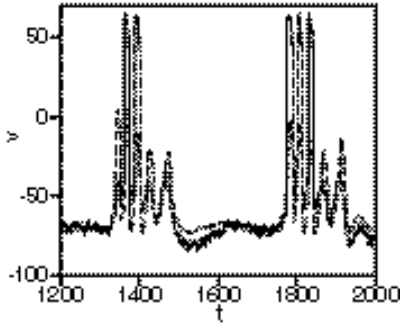}
\end{center}
\caption{Time evolutions of $v_i$ (solid line) and the average $\langle v\rangle$ (dashed line) in the Langevin simulation corresponding to Eq.~(8) with $\tau_1=5,\,\tau_2=\tau_3=100,\,K_1=K_3=-30,\,K_2=30,\,D=0.2, I_{01}=-0.25,I_{02}=1$, and $N=1000$.}
\label{fig:7}
\end{figure}
The spindle oscillation is also seen in the corresponding Langevin simulation.
Figure 7 displays the time evolutions of the membrane potential $v_1$ of the first RE neuron and the average value $\langle v\rangle$ for the RE neurons at $\tau_1=5,\,\tau_2=\tau_3=100,\,K_1=K_3=-30,\,K_2=30,\,D=0.2,\, I_{01}=-0.25,\,I_{02}=1$ and $N=1000$, which corresponds to Fig.~6(b). The time evolution of the average value  $\langle v\rangle$ exhibits the doubly periodic oscillation  similar to Fig.~6(b). The firing of each neuron is well synchronized to the averaged membrane potential, but the firing is sometimes skipped and the intermittent firing is seen. The intermittent firing might be characteristic of strongly inhibited systems. Similar intermittent firing of the RE neurons is observed experimentally~\cite{rf:27}.

\begin{figure}[htb]
\begin{center}
\includegraphics[height=4.5cm]{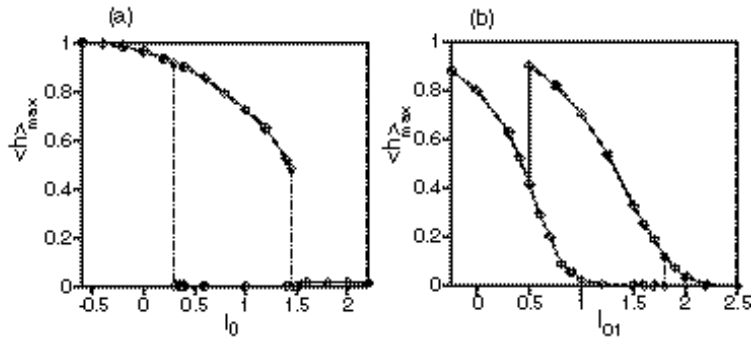}
\end{center}
\caption{Maximum value of the average $\langle h_1(t)\rangle=\int_{-\infty}^{\infty}\int_0^1P_1(x,y,t)ydxdy$ as a function of (a) $I_0=I_{01}=I_{02}$, and (b) as a function of $I_{01}$ at $I_{02}=1$. Both figures display the existence of the bistability and the hysteresis.}
\label{fig:8}
\end{figure}
The bistability is also observed in this coupled system. 
Figure 8(a) displays the maximum value in the time evolution of the average $\langle h_1(t)\rangle=\int_{-\infty}^{\infty}\int_0^1P_1(x,y,t)ydxdy$ as a function of $I_0=I_{01}=I_{02}$. The upper branch represents the burst mode and the lower branch represents the tonic mode for the RE neurons. 
The bistability occurs in $0.3\le I_0\le 1.4$.    
Figure 8(b) displays the maximum value in the time evolution of the average $\langle h_1(t)\rangle=\int_{-\infty}^{\infty}\int_0^1P_1(x,y,t)ydxdy$ as a function of $I_{01}$ for a fixed value of $I_{02}=1$. The bistability is observed in $0.5\le I_{01}\le 1.8$, although the classification to the two burst and tonic modes seems to be difficult in this case.
The thalamus is considered to be a relay point between the sensory organs and the cerebrum. The bistability of the activity of the coupled system of the RE cells and TC cells might be related to some switching function of the information transmission at the relay point. 

\section{Summary}
We have proposed another IF model including the dynamics in the excited state. We have studied the stability of the synchronized motion of a coupled system of two neurons and $N$ neurons.  In our model, the synchronization is rather robust  even in the case of inhibitory coupling. 

We have investigated global synchronization of noisy IF neurons with inhibitory coupling using the Fokker-Planck equation, and found the reentrant Hopf bifurcation. In the parameter region of weak coupling, almost all neurons fire synchronously, but in the parameter region of strong coupling, only a fraction of neurons fires synchronously owing to the strong inhibition. Intermittent firing is seen in each neuron as is clearly shown by the Langevin simulation. 
We have proposed another IFB model including the dynamics of the low-threshold Ca$^{2+}$ current based on the new IF model. We have performed direct numerical simulations of the Fokker-Planck equation for the globally coupled IFB model, and found the bistability of the tonic model and the burst mode. 
In the burst mode, the low-threshold Ca$^{2+}$ current facilitates the global oscillation.
We have further performed numerical simulations of the Fokker-Planck equation for the coupled system of the RE neurons and TC neurons, and found a doubly-periodic motion similar to the spindle oscillation observed in the second stage of the sleep. The doubly-periodic motions are also bistable in a certain parameter range of the inputs, which might be related to a switching function of information transmission at the relay point. 

We have found various bifurcations including the Hopf bifurcation, period doubling bifurcation and jump phenomena between bistable states in globally coupled noisy IF and IFB models.

Our model includes many control parameters. Our choice of the parameters may not be physiologically suitable. In future work, we will adjust the parameter values  and perform numerical simulations more relevant to realistic neurons.

\end{document}